\documentstyle[aps,prl,epsf,floats]{revtex}

\begin{document}
\draft
\preprint{}
\twocolumn[\hsize\textwidth\columnwidth\hsize\csname@twocolumnfalse%
\endcsname

\title{Spectral properties of statistical mechanics models}

\author{Hendrik Meyer, Jean-Christian Angl\`es d'Auriac,
 and Henrik Bruus}
\address{
Centre de Recherches sur les Tr\`es Basses Temp\'eratures,
BP 166, 38042 Grenoble, France}

\date{April 5, 1996}

\maketitle

\begin{abstract}
The full spectrum of  transfer matrices of the general
eight-vertex model on a square lattice is obtained by numerical
diagonalization. The eigenvalue spacing distribution
and the spectral rigidity are analyzed. 
In non-integrable regimes we have found eigenvalue repulsion
as for the Gaussian orthogonal ensemble in random matrix theory.
By contrast, in integrable regimes we have found
eigenvalue independence leading to a Poissonian behavior,
and, for some points, level clustering.
These first examples from classical statistical mechanics
suggest that the conjecture of integrability successfully
applied to quantum spin systems also holds for classical systems.
\end{abstract}

\pacs{PACS numbers:  05.50.+q, 05.20.-y, 05.45+b}
]

Since the work of Wigner \cite{Wigner}
random matrix theory (RMT) has been applied successfully in various
domains of physics~\cite{Mehta}.
Recently several quantum spin Hamiltonians have been investigated
from this point of view.
It has been found \cite{Montambaux,Hsu} that 1D systems for which the Bethe
ansatz applies have a level spacing distribution
close to a Poissonian (exponential) distribution, $P(s) = \exp(-s)$,
whereas if the Bethe ansatz does not apply, the level spacing distribution
is described by the Wigner surmise for the Gaussian orthogonal ensemble (GOE),
$P(s) = \frac{\pi}{2} s \exp( -\pi s^2 / 4)$.
Similar results have been found for 2D systems \cite{Bruus}.
This suggests that the GOE describes
properly {\em some} properties of the spectrum of complex
quantum systems.
In this letter we extend the RMT analysis from quantum spin systems
to models of classical statistical mechanics. In particular,
we look at possible consequences of integrability
on the spectral properties of the transfer matrices.

At first sight it seems natural to start with the Ising model in
two dimensions without magnetic field as an example of an integrable model, 
and then to add a magnetic field.
It turns out that the spectrum of transfer matrices of the 
Ising model for finite size leads to numerical difficulties as
explained below. We then have chosen the
case of the general eight-vertex model on a square lattice (which
contains the zero-field Ising model as a special case) \cite{BaBook}. 
Moreover, it is known that zero-field eight-vertex transfer matrices
commute with the Hamiltonian of the anisotropic XYZ quantum spin chain
for certain relations between the parameters of the two models 
\cite{commute}.
We shall use the notation of Ref.~\cite{BaBook} to designate
the eight admissible vertices and their respective Boltzmann weights
$a$, $a'$, $b$, $b'$, $c$, $c'$, $d$, and $d'$.
We consider the row-to-row transfer matrices $T_N$ to
build iteratively a periodic rectangular lattice
by adding rows of length $N$ with periodic boundary conditions.
Therefore
the partition function of a periodic rectangle of $n$ rows of $N$ sites
is $Z_{n,N} = {\rm Tr} (T_N^n)$.
Note that there can be different expressions for the matrix $T_N$,
but all these expressions have the same value of the trace of the
$n$th power for any $n$, and therefore they can be deduced from each other
by a similarity transformation; the spectrum is indeed an intrinsic
quantity which does not depend on any particular choice
of the transfer matrix. To perform the usual statistical analysis of the
spectrum we need to have real eigenvalues. However, in general this is not 
the case and we will restrict ourselves to cases where 
the transfer matrix is symmetric. It is well known that the eight-vertex
model can be mapped onto an anisotropic Ising model on a square lattice with 
diagonal interactions and four spin interactions around each
plaquette. We again use the notations of Ref.~\cite{BaBook}
and introduce the five coupling constants 
$J_h$, $J_v$, $J$, $J^\prime$ and $J^{\prime\prime}$.
The transfer matrix of the spin model 
can be chosen symmetric if $J = J^\prime$. In terms
of the Boltzmann weights of the transfer matrix it requires
only that $c=d=c^\prime=d^\prime$ 
($c=c^\prime$ and $d=d^\prime$ is not a restriction). 
This condition is verified for
models without electrical field (i.e.\ when $a=a^\prime$ and
$b=b^\prime$) and also for models with a field.
So we are able to build symmetric transfer matrices
for integrable cases with $a=a^\prime$, $b=b^\prime$, and
$c=c^\prime=d=d^\prime$, as well as for non-integrable cases
with $a \neq a^\prime$, $b \neq b^\prime$, and
$c=c^\prime=d=d^\prime$.

Before presenting our results, we briefly recall some
features of the RMT analysis, which is a statistical analysis of
the eigenvalues of a given matrix regarded as an ordered set.
Firstly, one has to sort
the eigenvalues according to the symmetry of the corresponding eigenstate.
In contrast to quantum spin systems, transfer matrices posses
a priori no continuous symmetry (as e.g.\ the SU(2) spin symmetry),
but only space symmetries. For row-to-row transfer matrices these
are given by the automorphy group of a single row (and not of the full
lattice). This group is the set of permutations $g$ of sites such that
$g(i)$ and $g(j)$ are neighbours if and only if $i$ and $j$ are
neighbours. To each $g$ acting on the set of vertices
one can easily associate a linear operator $\hat g$ acting in the
configuration space. Obvioulsy $\hat g$ and $T_N$ commute. 
It is then possible to
construct a set of projectors onto invariant subspaces of $T_N$.
This amounts to block-diagonalize $T_N$. This is not only a useful way of
lowering the size of the matrices to diagonalize, but also the manner
to sort the eigenvalues. The automorphy group of
the periodic ring of length $N$ is the dihedral group $D_N$ 
generated by a translation and a reflection.
Elemenentary group theoretical analysis can be performed to build
the ($N/2+3$ if $N$ is even, or $(N-1)/2 +2$ if $N$ is odd)
projectors onto the invariant subspaces.
The transfer matrix of the zero-field model is also invariant under
the reversal of all arrows of the vertices and another projector has to
be applied.
Secondly,
to find universal behavior within each invariant subspace,
one needs to `rescale' the eigenvalues $E_i$
in order to have a {\em local} density of
eigenvalues equal to one. This operation is called the
``unfolding'' and produces the unfolded eigenvalues $\epsilon_i$.
The aim is to remove the non-universal or system specific
large scale variations of the integrated density of states,
and to study only the presumably universal short scale fluctuations.
It amounts to compute
an average integrated density of states $N_{av}(E)$ which is the
smooth part of the actual integrated density of states. We then have
$\epsilon_i=N_{av}(E_i)$.
In the generic case, several methods can be used to compute $N_{av}(E)$:
running average unfolding
(local averaging of eigenvalues followed by interpolation),
Gaussian unfolding (Gaussian broadening of each delta peak in the
density of states), and
Fourrier unfolding (removal of short scale wave lengths using
Fourrier transformation).

The simplest quantity one calculates in RMT analysis is the
distribution $P(s)$ of the differences between two consecutive
unfolded eigenvalues $s_i=\epsilon_{i+1}-\epsilon_{i}$. For
integrable systems a Poissonian distribution is expected, since the
$\epsilon_i$ are expected to be independent. In contrast, for the simplest
non-integrable systems the Wigner surmise is expected.
Another quantity of interest is the spectral rigidity \cite{Mehta}:
$$
\Delta_3(L) = 
\left\langle \frac{1}{L} \min_{a,b}
\int_{\alpha-L/2}^{\alpha+L/2}
{\left( N(\epsilon)-a \epsilon -b\right)^2 d\epsilon} \right\rangle_\alpha \;,
$$
where $\langle\dots\rangle_\alpha$ denotes an average over $\alpha$.
This quantity measures the deviation from equal spacing.
For a totally rigid spectrum, as that of the harmonic oscillator, one has
$\Delta_3^{\rm osc}(L) = 1/12$, for an integrable (Poissonian) system one has
$\Delta_3^{\rm Poi}(L) = L/15$, while for the Gaussian Orthogonal
Ensemble one has $\Delta_3^{\rm GOE}(L) = \frac{1}{\pi^2} (\log(L) -
0.0687) + {\cal O}(L^{-1})$. 
It has been found that the spectral rigidity of  quantum spin systems 
follows $\Delta_3^{\rm Poi}(L)$ in the integrable case and
$\Delta_3^{\rm{GOE}}(L)$ in the non-integrable case.
However, in both cases, even though $P(s)$ is in good agreement with RMT,
 deviations from RMT occur for $\Delta_3(L)$ at some system dependent
point $L^*$.
This stems from the fact that the rigidity
$\Delta_3(L)$ probes correlations beyond nearest neighbours
 in contrast to $P(s)$.
This is probably why the rigidity is much more sensitive
to the parameters of the unfolding than the spacing distribution. 

\begin{figure}[htb]
\epsfxsize=\columnwidth\epsfbox{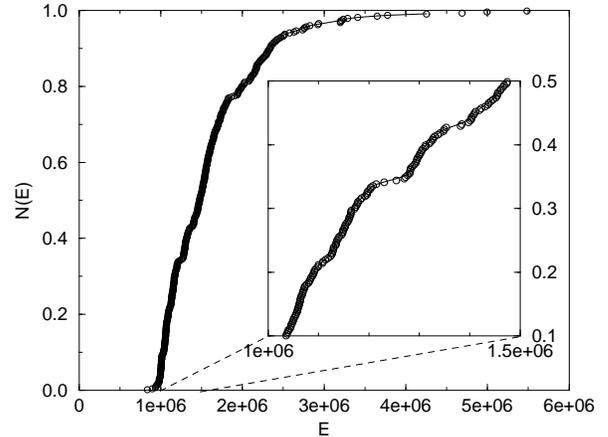}
\caption{The integrated density of eigenvalues $N(E)$ (circles) for the case
$a=2.5$, $a^\prime=1.6$, $b=b^\prime=3$, $c=c^\prime=d=d^\prime=1/
\protect\sqrt 6$ for a single symmetry invariant block. The insert
shows how complex $N(E)$ is even at a very fine scale.
The full line is $N_{av}(E)$.
}
\label{DOS}
\end{figure}

\begin{figure}[htb]
\epsfxsize=\columnwidth\epsfbox{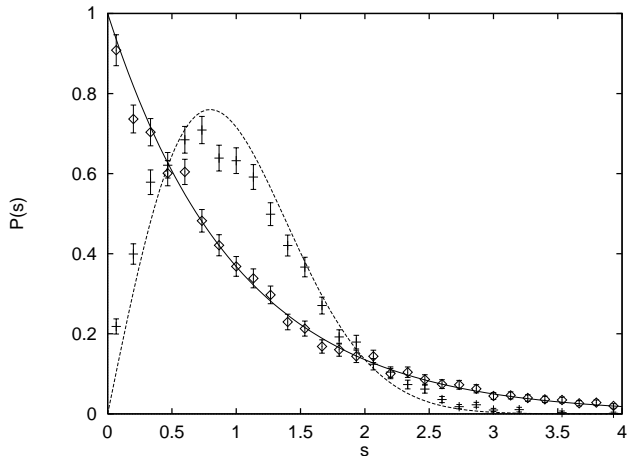}
\caption{ The eigenvalue spacing distribution $P(s)$ for
two row-to-row transfer matrices of the eight-vertex model with $N=14$.
The diamonds correspond to 
$a=a^\prime=2$, $b=b^\prime=3$, $c=c^\prime=d=d^\prime=1/
\protect\sqrt 6$:
this corresponds to an integrable point in the ordered
region of the phase diagram without electrical field.
The plus signs correspond to
$a=2.5$, $a^\prime=1.6$, $b=b^\prime=3$, $c=c^\prime=d=d^\prime=1/
\protect\sqrt 6$.
The full line is the Poissonian distribution while the dashed line is
the Wigner surmise.}
\label{PDSPoisGOE}
\end{figure}

\begin{figure}[htb]
\epsfxsize=\columnwidth\epsfbox{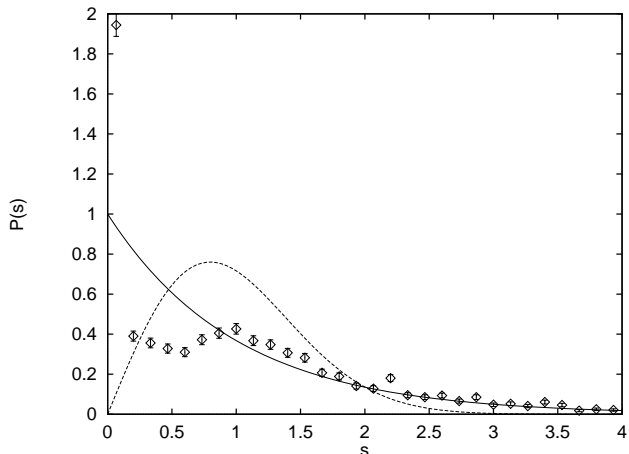}
\caption{The eigenvalue spacing distribution $P(s)$ for
a point satisfying the free-fermion condition.
The Boltzmann weights are
$a=0.8$, $a^\prime=1/a$, $b=b^\prime=\protect\sqrt{2c^2-1}$, and
$c=c^\prime=d=d^\prime=1/1.1$.
Note the peak near $s=0$ and the Poissonian tail for $s>1$.}
\label{PDSFF}
\end{figure}

We have generated  transfer matrices for different values of the
Boltzmann weights and linear size $N=16$ (resp. $N=14$). This
leads to matrix sizes of $65536^2$ (resp.\ $16384^2$).
Because of the eight-vertex condition, the transfer matrix couples only
configurations with the numbers of up (or down) arrows having the same
parity, so the matrix trivially seperates into two blocks.
After projection 
the matrix splits up into 18 (resp.\ 16) symmetry invariant blocks
of which the largest has a size of $2062^2$ (resp.\ $594^2$).
We have tried several methods of unfolding.
The results presented here are  obtained using either a Gaussian
unfolding with a local broadening over five states, 
or a running average unfolding over ten states.
In Fig.~\ref{DOS} we present
a typical integrated density of eigenvalues together
with the averaged curve. The spectrum is seen to be rather more complex
than a usual spectrum of a quantum spin system, thus the unfolding has to
be performed very carefully.
Fig.~\ref{PDSPoisGOE} shows the
probability distribution of the eigenvalue spacings averaged over all
representations for two representative sets of Boltzmann weights.
The diamonds correspond to a zero-field
case with $a=a'=2$, $b=b'=3$, and $c=c'=d=d'=1/\sqrt 6$. 
This point lies in an ordered region of the phase diagram, and
the transfer matrix has eigenvectors of the Bethe ansatz form.
The spacing distribution is close to the Poissonian distribution.
The plus signs on the same figure correspond to the case
$a=2.5$, $a^\prime=1.6$, $b=b^\prime=3$ $c=c^\prime=d=d^\prime=1/\sqrt 6$.
This point is in a region of parameter space where the Bethe ansatz
does not apply.
The spacing distribution is close to the Wigner surmise.
In particular the level repulsion is clearly seen.
To test further the idea that integrability leads to a Poisson law
and non-integrability leads to the Wigner surmise we also 
have calculated the spectrum of transfer matrices for some particular
points satisfying the free-fermion condition $aa'+bb'=cc'+dd'$ \cite{FanWu}.
In this case the results are less clear.
In Fig.~\ref{PDSFF} we have chosen a free-fermion point
within the generally non-integrable region where an electrical field
is present.
The Boltzmann weights are
$a=0.8$, $a^\prime=1/a$, $b=b^\prime=\protect\sqrt{2c^2-1}$, and
$c=c^\prime=d=d^\prime=1/1.1$.
The spacing distribution is close neither to the Poissonian distribution
nor to the Wigner surmise. In particular there is a 
peak at $s=0$ indicating level clustering.
We have found the same phenomenon for the pure 2D Ising 
model which can be mapped onto a eight-vertex model
satisfying the free-fermion condition. This behavior is usually found
together with a very involved density of eigenvalues and leads to
numerical difficulties in the unfolding.
A possible explanation of the peak at $s=0$ could be
that the free-fermion model is a trivial non-generic
model as for example the Hubbard model at zero Coulomb
repulsion.
However, for some other values of the Boltzmann weights
also obeying the free-fermion condition, the spacing distribution 
is much closer to the Poissonian distribution. This suggests as another 
possible explanation the existence of quasi-degeneracy leading 
to a Shnirelman peak at the origin \cite{Chirikov}
for this specific set of Boltzmann weights.
This will be studied in detail in a forthcoming publication.
From the above results we conjecture that 
the spacing distribution of eigenvalues of non-integrable models
is close to the Wigner surmise corresponding to level
repulsion, while for integrable models there is no level repulsion.
In  integrable systems there is level independence
leading to a Poissonian spacing distribution, but with a tendency 
to level attraction in some cases.

\begin{figure}[htb]
\epsfxsize=\columnwidth\epsfbox{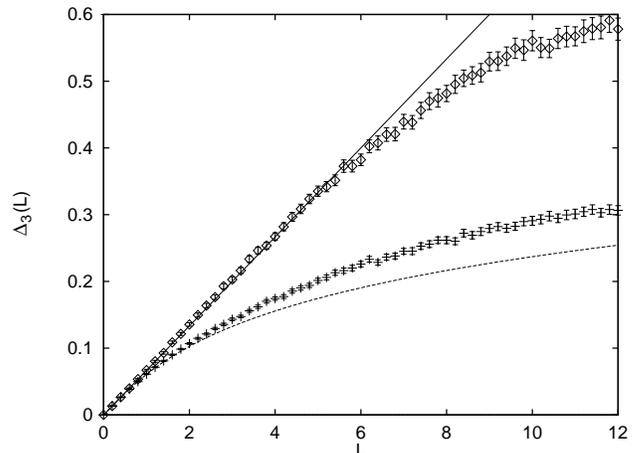}
\caption{The spectral rigidity $\Delta_3(L)$ for the same cases
as in Fig.~\protect\ref{PDSPoisGOE}.}
\label{D3PoisGOE}
\end{figure}

To go further in the RMT analysis, we present in Fig.~\ref{D3PoisGOE}
the spectral rigidity $\Delta_3(L)$
for the same points in parameter space corresponding to integrability
and to non-integrability as in Fig.~\ref{PDSPoisGOE}. The two
limiting cases corresponding to the Poissonian distributed
eigenvalues (solid line) and to GOE distributed eigenvalues (dashed line)
are also shown.
For the integrable point the agreement between the numerical
data and the expected rigidity is good up to a value $L\approx5$.
For larger values of $L$
a saturation occurs showing the limitation of the model
of independent eigenvalues. For the non-integrable case
the departure of the rigidity from the expected behavior
appears at $L \approx 2$, indicating 
that the RMT is only valid at short scales. Such behavior
has already been seen in quantum spin systems \cite{Bruus}.
We stress that these numerical results depend much more on the
unfolding than the results concerning the spacing distribution.

In summary, we have numerically calculated the spectrum of
transfer matrices of the 2D eight-vertex model for various parameters.
After having sorted and unfolded the spectrum we have computed
the eigenvalue spacing distribution and the spectral rigidity
averaged over all representations. To our knowledge
this is the first RMT analysis of  transfer matrices of 
models in classical statistical mechanics.
We have found that the non-integrable
cases are well described at short scales by a the Gaussian orthogonal
ensemble, while in the integrable cases the eigenvalues are mostly
independent. We speculate that this is a general results.

\acknowledgments
We would like to thank
J.M.\ Maillard for many discussions concerning integrability
of vertex models.
H.B.\ is supported by the European Commission under grant no.\ ERBFMBICT
950414.

\end{document}